\documentclass[aps,prl,twocolumn,nopacs,superscriptaddress]{revtex4-1} 

\usepackage{amsmath}
\usepackage{bbm}
\usepackage{epsfig}
\usepackage{epstopdf}
\usepackage{color}

\newcommand{\id}{\mathbbm{1}}

\usepackage{times}

\newcommand{\rr}{{\mathbbm{R}}}
\newcommand{\zz}{{\mathbbm{Z}}}
\newcommand{\nn}{{\mathbbm{N}}}

\newtheorem{theorem}{Theorem}

\newtheorem{observation}[theorem]{Observation}

\begin{document}

\title{Positive Wigner functions render classical simulation of quantum computation efficient}

\author{A.\ Mari} 
\affiliation{Dahlem Center for Complex Quantum Systems, Freie Universit{\"a}t Berlin, 
14195 Berlin, Germany}
\affiliation{Institute for Physics and Astronomy, University of Potsdam, 
14476 Potsdam, Germany}
\affiliation{NEST, Scuola Normale Superiore and Istituto di Nanoscienze - CNR, Pisa, Italy}

\author{J.\ Eisert}
\affiliation{Dahlem Center for Complex Quantum Systems, Freie Universit{\"a}t Berlin, 
14195 Berlin, Germany}

\begin{abstract}
We show that quantum circuits where the initial state and all the following quantum operations
can be represented by positive Wigner functions can be classically efficiently simulated.
This is true both for continuous-variable as well as discrete variable systems in odd prime dimensions,
two cases which will be treated on entirely the same footing.
Noting the fact that Clifford and Gaussian operations preserve the positivity of the Wigner
function, our result generalizes the Gottesman-Knill theorem. Our algorithm provides
a way of sampling from the output distribution of a computation or a simulation,
including the efficient sampling from an approximate output distribution in case of 
sampling imperfections for initial states, gates, or measurements. In this sense, this work highlights the role of the positive Wigner 
function as separating classically efficiently simulatable systems from those that are 
potentially universal for quantum computing and simulation, and it emphasizes the role of negativity of the 
Wigner function as a computational resource.
\end{abstract}

\maketitle

What renders a quantum computer a superior computational device? Where is the precise boundary between classically
efficiently simulable problems and ones for which this is no longer possible? Despite a significant research effort and partial
progress \cite{Nielsen,MaartenResources, MaartenSimulation,VanDenNest,VidalSimulation, ShiSimulation, FCS,TooMuch,Gottesman,Campbell}, 
these questions are still largely open. Quantum correlations surely play a role in one way or the other in 
quantum computers and simulators outperforming their classical counterpart. For example, if the 
entanglement is -- in a precise sense -- 
too low in a pure-state computation with respect to any bi-partite split, then one can classically
efficiently simulate the dynamics \cite{MaartenResources, MaartenSimulation,VidalSimulation, ShiSimulation, FCS}. 
In measurement-based computing specifically \cite{Oneway,MBQC}, 
where the resource character of entanglement is particularly manifest, states can be too little
entangled \cite{MaartenResources}, but also in a sense too entangled \cite{TooMuch}. 
Possibly the singularly most important result
on classical simulatability of quantum computers is the Gottesman-Knill theorem, stating that stabilizer circuits
consisting of Clifford gates only can be classically efficiently simulated \cite{Nielsen,Gottesman,GraphStates}. 
Though, adding almost any further gate will render the Clifford gate set universal for quantum computing. Similarly,
Gaussian operations for continuous-variable systems can be efficiently simulated \cite{Sanders,NoGo}.

Here we present a generalization of the Gottesman-Knill theorem, stating that one can efficiently classically
sample from quantum circuits starting from product states
with a positive Wigner function, applying quantum gates that have a positive Wigner function (in a sense
made precise below) and performing measurements associated with positive Wigner functions. 
This result holds true both for discrete variable systems where the
constituents have odd prime dimension (it can easily be generalized to arbitrary odd dimension) 
as well as for continuous-variable systems so common in quantum optics.
In fact, these two situations can be treated on exactly the same footing -- since at the root of the remarkably
simple argument, we exploit the structure of the phase spaces, discrete as well as continuous. The relative
elegance of the approach draws from the observation that the expressions for the discrete and continuous
description are identical.

In this sense, negativity of the Wigner function is identified as a computational resource. Indeed, it 
is usually seen as an indicator of ``non-classicality'' \cite{Direct,Positive,Galvao} 
(implying the $P$-function to be non-positive). 
Here we see that it is exactly this negativity that is needed if a quantum computer or simulator
is to outperform its classical counterpart. This results adds meat to the notion of such states being ``classical'',
quite similar to the situation of states with positive Wigner functions and homodyne measurements being unable to
violate a Bell inequality \cite{Bell} or be useful in magic state distillation \cite{Magic}. 
We also comment on the converse direction, that quantum computing
is possible with resources having Wigner functions arbitrarily close to being positive. 
While negativity is a necessary 
resource, one does not need to have ``a lot of it'' and it can be used in a dilute form. 

\section{ Discrete and continuous phase spaces}

We start by discussing the less commonly 
addressed phase space structure of finite-dimensional
quantum systems (compare Refs.\ \cite{PosWig,Margulis,Vourdas}), 
and shift the emphasis to infinite-dimensional ones later.
We assume the local dimension $d$ to be an odd prime, 
merely to avoid technicalities required in non-prime dimension (but they can be
treated on the same footing \cite{PosWig}).  
In this case the {\it phase space} of a single $d$-level systems 
is $\zz_d\times \zz_d$, so that it can be 
associated with a $d\times d$ cubic lattice.
Given an orthonormal basis $\{|0\rangle, |1\rangle, \dots |d-1 \rangle \}$, 
one can define the {\it shift} and {\it boost operators} 
as the generalizations of the familiar Pauli matrices by
\begin{eqnarray}\label{shiftboost}
  x(q)| x \rangle= |x+q \mod d\rangle, \quad\quad
	z(p)| x \rangle = \omega^{p x} | x\rangle,
\end{eqnarray}
where $\omega=e^{2\pi i/d }$ is
a $d$-th root of unity, arithmetic being modulo $d$, and $x=0,\dots, d-1$. 
The fundamental tools of every quantum phase space representation are the so called 
{\it Weyl operators}. For discrete systems they are given by
\begin{equation}\label{eqn:weyl}
	 w(q,p)=\omega^{-2^{-1}p q} z(p)x(q),
\end{equation}
where $2^{-1}=(d+1)/2$ is the multiplicative inverse of $2$ (again modulo $d$).
These operators form a group, the Heisenberg-Weyl group, and are the main ingredient
for representing quantum systems in phase space.

Let us now turn to continuous systems, and focus on a single mode first, associated
with canonical coordinates or ``quadratures'' of position $Q$ and momentum $P$. 
The associated phase space is now $\mathbbm R^2$. Indeed,
the continuous version of the Weyl operator can be analogously given in terms of the previous formulas
(\ref{shiftboost}) and (\ref{eqn:weyl}), with the only difference using standard arithmetic.
In fact, with a different choice of the phase factor, namely $\omega=e^{i}$, the Weyl operator in
Eq.\ (\ref{eqn:weyl}) takes the form 
\begin{equation}
	 w(q,p)=\exp(i q P - i p X),
\end{equation}
which is the familiar displacement operator well known in quantum optics.
From now on we are going to use a unique notation defined in such a way to be consistent with
discrete and continuous phase space representations. 
For this reason, in Table \ref{symb} we introduce a set of symbols which are valid in both settings. 

\begin{table}[h!]
\begin{tabular}{c|c|c} 
         Symbol & Discrete & Continuous \\ \hline 
           $\omega$  & $e^{2\pi i/d }\; \mod \; d$ & $e^i$  \\
          $w(q,p)$     & $\omega^{-2^{-1}p q} z(p)x(q)$ & $\exp(i q P - i p X)$ \\
              $r$      &     $(q_1,p_1,\dots, q_n,p_n)\in \mathbbm Z_d^{2n}    $   &$(q_1,p_1,\dots, q_n,p_n)\in \mathbbm R^{2n}    $  \\
             $\oint_r$ &$ \sum_{r\in \mathbbm Z_d^{{\rm size}(r)}}$& $ \int_{r\in \mathbbm R^{{\rm size}(r)}}dr^{{\rm size}(r)}$ \\
	      $c$      &      $2d$                    & $2\pi$  
\end{tabular} 
\caption{Table of symbols providing a unified notation valid for both discrete and continuous-variable
          systems. The function ${\rm size}(r)$ gives the number of constitutents of $r$, so that the
          sum (or the integral) is correctly defined also for local subsystems.} \label{symb}
\end{table}

The Wigner function of an operator $O$ acting on $n$ discrete or continuous-variable systems
is defined as 
\begin{equation}
	W_{O}(r)=(c/2)^{-n}{\rm tr} \left(w(r) \Pi^{\otimes n} w(r)^\dag O\right).
\end{equation}
where $w(r) = w(q_1,p_1)\otimes \dots\otimes w(q_n,p_n)$ and $\Pi$ is the single system parity 
operator: For discrete systems, on state vectors, this parity operator acts as
	$\Pi: | x \rangle \mapsto |-x \mod d\rangle$
for continuous-variable systems as
	$(\Pi\psi)(x) = \psi(-x)$.
%
%
This function has the structure of $W:\zz_d^{2n}\rightarrow \rr$ for discrete systems (compare also
Ref.\ \cite{Positive})
and $W:\rr^{2n} \rightarrow \rr$ for continuous systems.

\subsection{Properties of the Wigner function}
The Wigner function of a state $\rho$ is normalized and can be interpreted as the quantum 
analogue of a phase space distribution with the peculiar property of being not necessarily positive
in all its domain.
In the next sections we are going to use three important properties of the Wigner function:
\begin{eqnarray}
 {\rm tr}(\rho) &=&\oint_r W_\rho(r)=1,  \label{norm1} \\
 {\rm tr}(A B) &=&c^{2n} \oint_r W_A(r)W_B(r), \label{parseval}\\
 W_{A^T}(r)&=&   W_{A}(\Lambda r), \quad \Lambda={\rm diag}(1,-1,...,1,-1).
\end{eqnarray}

By virtue of Hudson's theorem, the only pure states having a positive Wigner function are
 {\it Gaussian states} \cite{CV} for continuous systems \cite{Hudson} and {\it stabilizer states} for 
odd local dimension $d$ \cite{PosWig}.
Unitary operations preserving the Gaussian form of the Wigner function are {\it Clifford operations}
in discrete systems and {\it Gaussian operations} in continuous systems. 
Those operations admit, via the Jamiolkowski isomorphism, a description 
in terms of positive Gaussian Wigner functions. 

For mixed states the situation is more complex. Surely, convex combinations of Gaussian and stabilizer
states will have positive Wigner functions. But there also exist states
with a positive Wigner function which cannot be represented in this form \cite{Werner, PosWig}. 
Analogously there are quantum operations admitting a positive Wigner representation but
which are not a convex combination of Clifford/Gaussian unitary operations.
In this work we will focus on the simulation of quantum algorithms involving states, operations 
and measurements all described by positive Wigner functions.
This corresponds to a more general scenario of which the Gaussian setting is a particular case. In this 
sense our result can be viewed as an extension of the Gottesman-Knill theorem. Quite remarkably,
our method it is completely independent from any Gaussianity property. 


\subsection{Operations having positive Wigner functions}
We have already introduced the Wigner function of a state or more generally
of an operator. We still have to properly give a phase space description to
operations and measurements. We now define the {\it Choi matrix} $f$ of a completely positive map
$F$. We define
\begin{equation}
	f= (\id\otimes F) |\omega\rangle \langle \omega| 
\end{equation}
where $|\omega \rangle=(\sum_{j=1}^{d} |j ,j\rangle)^{\otimes n}$ for discrete systems, 
which is up to normalization the state vector of the maximally entangled state. 
For continuous systems, 
if the limit exists and is of trace class, we set
\begin{equation}
	f= \lim_{s \rightarrow 1} (\id\otimes F) |\omega_s\rangle \langle \omega_s| 
\end{equation}
where
$|\omega_s\rangle =(\sum_{j=1}^\infty s^j |j,j \rangle)^{\otimes n}$. The Choi matrix is positive, as a
consequence of $F$ being completely positive, and it is supported on a Hilbert space having the natural structure of 
$\mathcal{H} = \mathcal{H}^{\rm in} \otimes \mathcal{H}^{\rm out}$.
We denote the partially transposed matrix with respect to subsystem ``out'' 
with the symbol $f^{\Gamma}$. 
We will say that the completely positive map $F$ ``has a positive Wigner function'' if the Choi matrix $f$ has
a positive Wigner function $W_f$. It is easy to see that if $f$ has a positive Wigner function,
then the same is true for $f^{\Gamma}$ as well.
The application of an operation to a state $\rho^{\rm out}=F(\rho^{\rm in})$
is related to the Choi matrix via a partial transposition $\Gamma$ and a partial trace
\begin{equation}
	\rho^{\rm out}={\rm tr}_{\rm in} ( (\rho^{\rm in} \otimes \id)
	f^\Gamma ).
\end{equation}
In phase space this is reflected by
\begin{equation}\label{d1}
	W(r^{\rm out})=c^{2n}\oint_{r^{\rm in} } 
	 W_{f^\Gamma} (r^{\rm out};r^{\rm in})W(r^{\rm in}) ,
\end{equation}
where $r^{\rm in}, r^{\rm out} \in \zz_d^{2n}$. 
Trace-preserving operations are such ${\rm tr}_{\rm out} (f)= \id$, i.e. 
\begin{equation}\label{stochastic}
 	c^{2n} \oint_{r^{\rm out}} W_{f^\Gamma}(r^{\rm out};r^{\rm in}) =1 ,
\end{equation}
for all $ r^{\rm in}$.
This means that, if the function is positive, $c^{2n}W_{f^\Gamma}$ can be interpreted 
as a {\it classical stochastic matrix}. 
This will be a key property for the classical sampling algorithm. For discrete systems, the Wigner function 
associated with the identity
operation is given by the Kronecker delta
\begin{equation}\label{delta}
	W_{|\omega\rangle \langle \omega|^\Gamma}=c^{-2n} \delta(r^{\rm out};r^{\rm in}).
\end{equation}
For continuous systems, one can also consider Wigner functions associated with operations for which the Choi matrix is not
of trace class, such as when $F$ is the identity operation. In this case, we have in the sense of distributions
\begin{equation}\label{delta}
	W_{|\omega_s\rangle \langle \omega_s|^\Gamma}\rightarrow c^{-2n} \delta(r^{\rm out};r^{\rm in}).
\end{equation}

An important subcase of operations having a positive Wigner functions 
is constituted by Clifford/Gaussian unitaries $U_{S,d}$.
They map Weyl operators onto Weyl operators under conjugation, so that $U_{S,d} w(r) U^\dagger_{S,d}=w(Sr+d)$
is again a valid Weyl operator. Here, $S\in Sp(2n,\zz_{d})$ and $v\in \zz_d^{2n}$ for discrete systems
while $S\in Sp(2n,\rr)$ and $v\in \rr^{2n}$ for continuous systems. 
From Eq.\ (\ref{delta}), one can deduce that the associated Choi matrix 
has the following Wigner function representation for discrete systems  \cite{PosWig}
\begin{equation}\label{clifford}
 	W_{f_{S,d}^\Gamma}=c^{-2n} \delta(S r^{\rm out}+d;r^{\rm in}),
\end{equation}
and similarly for continuous systems in the sense of distributions.
Physically, Clifford and Gaussian operations are of utmost importance in the laboratory since they
can often be realized with simple experimental settings. 
Since Clifford/Gaussian operations and convex combinations have a positive
Wigner function, our result covers also this setting.

\subsection{Measurements with positive Wigner functions}
We finally have to give a phase space description of a general measurement acting on single tensor
factors. Given a general measurement defined by a set of positive operator valued measures
(POVM) $\{M_k\}$ satisfying $\sum_{k=1}^K M_k=\id$,
we associate to each operator $M_k$ the respective Wigner function $W_{M_k}:\zz_d^2\rightarrow \rr$
for discrete systems and $W_{M_k}:\rr^2\rightarrow \rr$ for continuous systems.
Given a state $\rho$ the probability of getting the measurement outcome $k$ on subsystem $l$ will be
$P(k)={\rm tr} ( \rho M_k^{(l)} )$ and hence, for a single system,
\begin{equation}
 	P(k)=c^{2}\oint_r W_\rho (r) W_{M_k}(r_l).
\end{equation}
Moreover, we also have
\begin{equation}
 	c^{2}\sum_{k=1}^K W_{M_k}(r_l)=1  \label{norm2}
\end{equation}
for all $r$ both in the discrete and the continuous setting,
which means that, if the functions $c^{2} W_{M_k}$ are positive, then they 
can be interpreted as probabilities in the variable $k$ for each value of $r$. 
This fact will be also crucial in the simulation algorithm.

\section{ Statement of the problem}
Having laid out the formalism to be employed, we are now in the position to precisely state the problem at hand:
We allow for general quantum circuits of the following form, again both in the continuous or discrete setting.

\begin{itemize}
\item We consider initial product states $\rho=\rho_1\otimes\dots \otimes \rho_n$.
\item To this initial state, a 
sequence of trace preserving quantum channels 
$F_t \circ \dots  F_2 \circ  F_1$ is applied (not necessarily unitary), 
each of them supported on at most $m$ subsystems at a time.
\item Finally, local measurements are performed 
on each individual subsystem defined by some local projective positive operator valued measures
$\{M_k\}$.
\end{itemize}
Here $n$ denotes the number of constituents, while $t$ is the depth of the circuit, so the number of 
gates or local operations applied. The classical simulation should scale polynomially with respect to
 these two parameters.
We remark that initial states can be mixed and the applied gates will in general be non-unitary.
This is probably the most general model of a non-adaptive quantum algorithm,
{\it i.e.} where the sequence of gates and measurements is fixed. Non-adaptivity has be chosen 
just for simplicity of the exposition but the algorithm can be easily extended to the adaptive case.

A run of the quantum algorithm will provide one list of outcomes $k_1,k_2, \dots k_n$,
one for each measured subsystem. 
The probability of a given list of outcomes to occur is
\begin{equation}
	P(k_1,\dots, k_n)={\rm tr} \left(
	(M_{k_1}\otimes M_{k_2}\dots M_{k_n})( F_t \circ \dots F_2 \circ F_1 )
	( \rho) \right).
	 \label{qc}
\end{equation}
Using quantum systems, one can hence sample from the distribution $P$.
Classically, the problem to be solved is again a sampling problem:
The quantum circuit can be classically simulated if there is a classical algorithm that is efficient
in $t$ and $n$ and provides, in each run, a list of outcomes $k_1,\dots , k_n$ drawn from (approximately)
the same  probabilities
of the quantum circuit given in Eq.\ (\ref{qc}). 
Note that we do not require to really compute the probabilities of all the possible outcomes (simulation in a stronger
sense), but we just want a classical algorithm to be efficient
in sampling from the distribution defined by the 
quantum circuit, in the sense that output strings
of the classical and quantum machines are drawn from 
the same (or approximately the same) probability distribution. This weaker sense of simulation
is anyway enough to exclude any possible speedup of the quantum algorithm with respect to the classical one,
and hence identifies negativity of the Wigner function as a necessary resource in quantum computing.
We will also take into account possible errors in the classical sampling, showing that 
those errors cannot strongly affect the final probability distribution. 

\subsection{Phase space representation}
First note that the 
Wigner function of the input product state $\rho=\rho_1\otimes\dots \otimes \rho_n$
is given by a product function $W_\rho$ with
\begin{equation}
	W_\rho(r)=W_{\rho_1}(r_1)W_{\rho_2}(r_2)\dots W_{\rho_n}(r_n),
\end{equation}
where $W_{\rho_l}$ is associated with the subsystem $\rho_l$. 
To each gate $F_t$ can be associated the Wigner function $W_{f_t^{\Gamma}}$.
The Wigner functions of the POVM  associated with the outcome $k$ and performed on subsystem $l$
are denoted as $W_{M_k}$ with local phase space coordinates $r_l$.
By sequentially applying Eq.\ (\ref{parseval}), we can express the outcome probabilities of the quantum circuit as
 \begin{eqnarray}
&& P(k_1,\dots, k_n)=\label{PWig}
c^{2n (t+1)}\oint_{r^{(t)},r^{(t-1)}\dots,r^{(0)}}   \\
&\times &W_{M_{k_1}}(r_1^{(t)})\dots W_{M_{k_n}}(r_n^{(t)})  \nonumber \\
&\times &W_{f_t^\Gamma}(r^{(t)};r^{(t-1)}) \dots W_{f_1^\Gamma}(r^{(1)};r^{(0)}) W_\rho(r^{(0)}),  \nonumber 
\end{eqnarray}
where the subscripts in the coordinates indicate subsystems while superscripts indicate the integer
time steps associated with the sequential application of the gates.
This is just a formal phase space description the quantum circuit completely equivalent to the operator
representation given in Eq.\ (\ref{qc}). Up to now no assumption has been made on the the type of resources
and operations. 

\section{Efficient classical simulation}

We now make the following two assumptions: 
\begin{itemize}
\item The Wigner functions of the input state, the gates, and the POVMs are positive,
\item it is possible to draw phase space points according to local probability distributions 
      associated with local states, local gates and local measurements. 
\end{itemize}
Later, the second of this assumptions will be relaxed allowing for eventual classical sampling errors.

\begin{observation}[Efficient classical simulation of circuits] For any $n$ and $t$ one can sample classically
from the distribution $P$ in $\text{poly}(n,t)$ time. 
\end{observation}

This is done as follows.
\begin{itemize}
\item {\it Step $0$:} Draw a phase space point $\bar r^{(0)}$ according to the input Wigner function $W_\rho(.)$, 
\item {\it Steps $j=1,\dots, t$:} draw a phase space point $\bar r^{(j)}$ according to the probability distribution 
$c^{2n}W_{f_{j}^\Gamma}(.; \bar r^{(j-1)})$.
\item {\it Step $t+1$:} Finally, draw a measurement outcome $k_1, k_2,\dots,k_n$ according 
to the probability distribution 
\begin{equation}
	P(k_1,\dots,k_n)=c^{2n}W_{M_{k_1}}(\bar r_1^{(t)})\dots W_{M_{k_n}}(\bar r_n^{(t)}).
\end{equation} 
\end{itemize}

{\it Proof:} This algorithm is simply a classical stochastic process and
directly from the law of conditional probabilities we have that, in the final step $t+1$, the 
probability of getting the outcome $k_1,k_2, \dots, k_n$ is given by Eq.\ (\ref{PWig}).
So one run of the classical algorithm is completely equivalent to one run of the quantum circuit.
Moreover, the efficiency of the classical procedure with respect to $n$ and $t$ follows from the following
observations:
\begin{itemize}
\item
Step $0$ is efficient because, since the initial state is a product, 
we merely have to draw $n$ independent subsystem phase space vectors. 
\item Steps $1,2,\dots,t$ are efficient because, for each gate in step $j$, 
we draw a phase space vector associated 
with at most $m$ local subsystems and leave the complementary coordinates invariant. 
This follows from the structure of $W_{f^\Gamma_j}$ factorizing as
(as a Kronecker delta or in the sense of distributions)
\begin{equation}
 	c^{2n}W_{f_{j}^\Gamma}(r^{j};r^{j-1})=c^{2m} 
	W^{(\rm local)}_{f_{j}^\Gamma}(r^{j}_L;r^{ j-1}_L) \delta(r^{ j}_C;r^{j-1}_C),
\end{equation}
where $r_L$ are the local coordinates of the $m$ subsystems involved in the gate, while $r_{C}$ are the complementary ones.
For every given input vector $ r^{ j-1}$ one has to draw a vector $ r^{ j}_L$ with respect to the local distribution
$c^{2m}W^{(\rm local)}_{f_{j}^\Gamma}$ and just leave the other
coordinates invariant $r^{j-1}_C \mapsto r^{j}_C$. 
\item Step $t+1$ is efficient because, since the final measurements are local, 
we draw $n$ independent outcomes associated to each subsystem.

\end{itemize}
As a final remark we observe that the positivity of all the Wigner functions and the properties given in Eqs.\ 
(\ref{norm1}), (\ref{stochastic}) and (\ref{norm2}) are crucial for all the functions 
appearing in the classical algorithm to be interpreted as probability distributions. If this is not the case, the classical
algorithm cannot be applied. However, using conditional probabilities, the algorithm can easily accommodate
adaptive later steps based on earlier measurement outcomes.

\section{Robustness to sampling errors}

In this section we discuss the robustness of the previous method with respect to possible 
errors in the classical sampling from positive Wigner functions.
Suppose that in the classical algorithm one is able to efficiently sample phase space
vectors only from imperfect probability distributions which are close to the 
ideal ones. These imperfect Wigner functions will correspond to imperfect density operators 
 $\rho'$, gates $F'$ with Jamiolkowski isomorphs $f'$ and POVM elements ${M_k'}$.
We will show that the classical algorithm is robust with respect to accumulated sampling errors. What is more,
these errors can be naturally linked to errors  in the physical quantum system.
For brevity, we focus on the discrete case here, while the continuous situation can be treated on the same footing.
For simplicity of notation, in turn, we denote the stochastic matrix associated with the $k$-th step
by 
$Q^{(k)}=c^{2n}W_{f_{k}^\Gamma}(.; .)$
and its (stochastic) approximation by $Q'^{(k)}$. $\|.\|_\infty,\|.\|_1 $ denote the usual respective 
matrix $p$-norms.

\begin{observation}[Efficient approximation of circuits] For any $n$, $t$, and any $\varepsilon>0$, 
consider a sampling from the distribution $P'$ obtained from initial states, operations, and measurements
deviating from those in Observation 1 in that
\begin{equation}
	|W_{\rho_l}(r_l) -W_{\rho'_l}(r_l)| <\varepsilon,\,
	|W_{M_{k_l}} (r_l)-W_{M'_{k_l}}(r_l)| <\varepsilon,\, 
\end{equation}
for $l=1,\dots, n$ and all $r_l$, and 
$\| Q^{(k)} - Q'^{(k)}\|_\infty <\varepsilon$ for $k=1,\dots, t$.
Then one can sample from $P'$ with $\|P-P'\|_\infty<\varepsilon \,\text{poly}(n,t)$ 
in $\text{poly}(n,t)$ time. 
\end{observation}

Note that the above estimates are also valid if one has trace-norm bounds for 
all states as well as operator norm bounds for the POVM elements.
Using the duality of the operator norm and the trace norm, one finds for all $r_l$
\begin{equation}
	 |W_{\rho_l}(r_l) - W_{\rho'_l}(r_l) | \leq   d^{-1} \| \rho_l - \rho'_l\|_1
\end{equation}
for $l=1,\dots, n$. Similarly, 
from the definition of the parity operator, we get
\begin{equation}
	|W_{M_{k_l}}(r_l) -  W_{M'_{k_l}}(r_l) |\leq \| M_{k_l} - M'_{k_l}\|_\infty
\end{equation}
for every $l=1,\dots, n$, any $r_l$, and each outcome.

{\it Proof:} Iteratively inserting and subtracting terms and using the triangle inequality 
as well as the fact that $|W_{\rho_l}(r_l)|\leq 1$ for all states $\rho_l$ and all $r_l$, we get for all $r$
\begin{equation}
	 | W_{\rho}(r)- W_{\rho'} (r) | \leq \sum_{l=1}^n |W_{\rho_l}(r_l) - W_{\rho'_l}(r_l) | \leq n  \varepsilon .
\end{equation}
Using the triangle
inequality several times, one finds
\begin{eqnarray}
	&&\sup_r \left|
	\biggl(\prod_{k=t}^1 Q^{(k)} W_\rho\biggr)(r) - \biggl(\prod_{k=t}^1 {Q'}^{(k)} W_{\rho'}\biggr)(r)
	\right| \\
	& \leq&
	\left\| 
	\prod_{k=t}^1 Q^{(k)}   - \prod_{k=t}^1 {Q'}^{(k)} 
	\right\|_\infty
	 +\sup_r
	\left| 
	\biggl(
	\prod_{k=t}^1 {Q'}^{(k)} (W_\rho -   W_{\rho'})\biggr)(r)
	\right| .\nonumber
\end{eqnarray}
Using the sub-multiplicativity of the operator norm, 
\begin{equation}
	\Bigl\| 
	\prod_{k=t}^1 {Q'}^{(k)}
	\Bigr\|_\infty \leq
	\prod_{k=1}^t \|{Q'}^{(k)}\|_\infty \leq1,\,\,  
\end{equation}	
as well as the fact that 
for all states $\rho$ and all $r$ we have that
$|W_{\rho}(r) |\leq 1$,
we find
\begin{eqnarray}
	&&\sup_r \left|
	\biggl(\prod_{k=t}^1 Q^{(k)} W_\rho\biggr)(r) - \biggl(\prod_{k=t}^1 {Q'}^{(k)} W_{\rho'}\biggr)(r)
	\right| \\
	& \leq& \sum_{k=1}^t
	\left\|  Q^{(k)} -  {Q'}^{(k)}
	\right\|_\infty+  n   \varepsilon\leq (t+n)\varepsilon
	. \nonumber
\end{eqnarray}
Using that
\begin{equation}
	\left| \prod_{l=1}^n
	W_{M_{k_l}}(r_l)-\prod_{l=1}^n
	W'_{M_{k_l}}(r_l)
	\right| \leq \sum_{l=1}^n |W_{M_{k_l}}(r_l)- W'_{M_{k_l}}(r_l)|
\end{equation}
and combining these results, one arrives at the above
observation.

%

\section{Summary and outlook}

At this point, one may ask ``how negative the Wigner function has to be'' in order to allow for universal quantum computing.
Surely, one can perform universal quantum computing with positive Wigner functions initial product 
states, as well as with circuits of quantum gates and local measurements, all of
which equipped with a Wigner function being arbitrarily close to being positive; so for each element of a
circuit, the answer is ``not much'':
For every $\varepsilon>0$ and $d=3$, there exist families of 
initial states $\rho_1,\dots, \rho_n$ with a positive Wigner function each, 
families of local gates with isomorphs $f_1,\dots, f_t$ supported on
a number of sites constant in $n$ and local POVMs $M_{k_1},\dots, M_{k_n}$ such that
\begin{equation}
	W_{f_j}(r)>-\varepsilon,\,\,
	W_{M_l}(r_l)>-\varepsilon,
\end{equation}
for $l=1,\dots, n$, $j=1,\dots, t$ and all $r_l,r$, giving rise to circuits
universal for quantum computing. This observation is obvious from the fact that 
product stabilizer measurements have a positive Wigner function. What is more, unitary gates can be
``diluted'' such that each gate is close to the identity operation. Also, by encoding the output, POVMs arbitrarily
close to the identity can be used. Similar schemes can also developed based on encoded cluster states
as in Refs.\ \cite{MBQC,BriegelEncoding}. Note that this is not in contradiction with Observation 2, since
despite the error growing slowly, it grows beyond all bounds in an unbounded 
computation using imperfect gates. 
A lesson to be learned from these rather obvious examples is that
it appears to be a very fruitful enterprise to meaningfully quantify the negativity as a resource in terms of a proper
resource theory, as it has happened in entanglement theory \cite{Brandao} or in statistical physics \cite{Oppenheim}.

In this work, we have shown that the negativity of the Wigner function can be grasped as a resource in quantum computing and
simulation: If the basic elements of a circuit exhibit a positive Wigner function, the probability distribution of the
quantum computation can be efficiently sampled. This remains true if one can only approximately implement each gate,
in that the errors made scale favorably. In this sense, our result generalizes the Gottesman-Knill theorem for sampling
outcomes of circuits, in a way where continuous and discrete systems are treated on exactly the same footing. 
We hope that the present approach 
stimulates further work on identifying the boundary between classically efficiently simulable quantum systems and
those universal for quantum computing.

\section*{Acknowledgements} 
This work has been supported by the EU (Q-Essence, MINOS), the BMBF (QuOReP), and the EURYI. 
We gratefully thank E.\ T.\ Campbell for helpful comments on the manuscript. 
Upon completion of this work, we became aware of Refs.\ \cite{New,New2}
which together make a similar claim on the efficient simulation of circuits having a positive Wigner function.

\section{Appendix}

\subsection{Preparation of random variables with given Wigner functions}
 
Here we briefly sketch how to draw
phase space points according to given Wigner functions, both for discrete as well as for continuous-variable systems.
There are several standard methods of how to proceed here, e.g., the rejection method 
or the inverse transform sampling method. We will be specific for 
sampling in $\rr^s$, $s\in \nn$, with a similar reasoning holding for $\zz_d^s$ as well.
Denote for a given non-negative Wigner function $W_\rho:\rr^s\rightarrow \rr$
with
\begin{equation}
	F: \rr^s\rightarrow \rr
\end{equation}
the cumulative distribution function, $r\in \rr^{2s}$. This can be written as
\begin{eqnarray}
	F(r_1,\dots, r_{2s}) &=& F_{R_1}(r_1) F_{R_2| R_1}(r_2|r_1)
	\\
	&\times &\dots
	F_{R_{2s}| R_1 \dots R_{2s-1}}
	(
	r_{2s}| r_1 \dots r_{2s-1}
	).\nonumber
\end{eqnarray}
Then, given $2d$ uniformly distributed random variables with realizations $u_1,\dots, u_{2s}$, 
take
\begin{eqnarray}
	r_1&=& F^{-1}_{R_1}(u_1),\\
	r_2&=& F^{-1}_{R_2{|} R_1}(u_2),\\
	&\vdots&\nonumber\\
	r_{2s}&=& F^{-1}_{R_{2s}}
	| R_1 \dots R_{2s-1}(u_{2s}).
\end{eqnarray}
This will then be distributed according to $W_\rho$.


\begin{thebibliography}{99}
		
\bibitem{Nielsen}
	M.\ A.\ Nielsen and I.\ Chuang, 
	{\it Quantum computation and quantum information}
	(Cambridge University Press, Cambridge, 2000).
	
\bibitem{MaartenResources}
	M.\ Van den Nest, A.\ Miyake, W.\ D{\"u}r, and H.\ J.\ Briegel,
	Phys.\ Rev.\ Lett.\ {\bf 97}, 150504 (2006). 

\bibitem{MaartenSimulation}
	M.\ Van den Nest, W.\ D{\"u}r, G.\ Vidal, and H.\ J.\ Briegel,
	Phys.\ Rev.\ A {\bf 75}, 012337 (2007).
	
\bibitem{VidalSimulation}
	G.\ Vidal,
	Phys.\ Rev.\ Lett.\ {\bf 98}, 070201 (2007). 

\bibitem{ShiSimulation}
	I.\ L.\ Markov and Y.\ Shi,
	SIAM J.\ Comp.\  {\bf 38}(3), 963 (2008).
	
\bibitem{FCS}		
	M.\ Fannes, B.\ Nachtergaele, and R.\ F.\ Werner,
	Commun.\ Math.\ Phys.\ {\bf 144}, 443 (1992).

\bibitem{TooMuch}
	D.\ Gross, S.\ T.\ Flammia, and J.\ Eisert, Phys.\ Rev.\ Lett.\ {\bf 102}, 190501 (2009).
				
\bibitem{Gottesman}
	D.\ Gottesman, 
	{\it Stabilizer codes and quantum error correction}, 
	PhD thesis (CalTech, Pasadena, 1997).

\bibitem{Campbell}
	E.\ T.\ Campbell and D.\ E.\ Browne, Phys.\ Rev.\ Lett.\ {\bf 104}, 030503 (2010).
	
\bibitem{VanDenNest}
	M.\ van den Nest, arXiv:1204.3107.

\bibitem{Oneway}	
	R.\ Raussendorf and H.\ J.\ Briegel, 
	Phys.\ Rev.\ Lett.\ {\bf 86}, 5188 (2001).	
	
\bibitem{MBQC}
	D.\ Gross and J.\ Eisert,	
	Phys.\ Rev.\ Lett.\ {\bf 98}, 
	220503 (2007).
		
\bibitem{GraphStates}
	For a discussion of the Gottesman-Knill theorem and efficient simulation of stabilizer circuits in terms
	of graph states \cite{Graphs}, see Ref.\ \cite{Briegel}.
	
\bibitem{Briegel}		
	S.\ Anders and H.\ J.\ Briegel, 
	Phys.\ Rev.\ A {\bf 73}, 022334 (2006).
		
\bibitem{Graphs}	
	M.\ Hein, J.\ Eisert, and H.\ J.\ Briegel, 
	Phys.\ Rev.\ A {\bf 69}, 062311 (2004).

\bibitem{Sanders}
	S.\ D.\ Bartlett, B.\ C.\ Sanders, S.\ L.\ Braunstein, and K.\ Nemoto,
	Phys.\ Rev.\ Lett.\ {\bf 88}, 097904 (2002);
	
\bibitem{NoGo}
	J.\ Eisert, S.\ Scheel, and M.\ B.\ Plenio, Phys.\ Rev.\ Lett.\ {\bf 89}, 
	137903 (2002); 
	J. Fiurasek, ibid.\ {\bf 89}, 137904 (2002);
	G.\ Giedke and J.\ I.\ Cirac, Phys.\ Rev.\ A {\bf 66},
	032316 (2002). 	

\bibitem{Direct}
	A.\ Mari, K.\ Kieling, B.\ Melholt Nielsen, E.\ S.\ Polzik, and J.\ Eisert, 
	Phys.\ Rev.\ Lett.\ {\bf 106}, 010403 (2011);
	   K.\ Laiho, K.\ N.\ Cassemiro, D.\ Gross, C.\ Silberhorn,
	   Phys.\ Rev.\ Lett.\ {\bf 105}, 253603 (2010).

\bibitem{Positive}
	C.\ Cormick, E.\ F.\ Galvao, D.\ Gottesman, J.\ Pablo Paz, and A.\ O.\ Pittenger,
	Phys.\ Rev.\ A {\bf 73}, 012301 (2006).

\bibitem{Galvao}
	E.\ F. Galvao,
	Phys.\ Rev.\ A {\bf 71}, 042302 (2005).
		
\bibitem{Bell}
	K.\ Banaszek and K.\ Wodkiewicz,
	   Phys.\ Rev.\ Lett.\ {\bf 82}, 2009 (1999).
	
\bibitem{Magic}
	V.\ Veitch, C.\ Ferrie, and
	J.\ Emerson, arXiv:1201.1256, version 1 of Jan.\ 5, 2012.

\bibitem{Margulis}
	D.\ Gross and J.\ Eisert, 
	Quant.\ Inf.\ Comp.\ {\bf 8}, 722 (2008).

\bibitem{Vourdas}
	S.\ Zhang and A.\ Vourdas,
	J.\ Phys.\ A {\bf 37},  8349 (2004).

\bibitem{Marchiolli}
	M.\ A.\ Marchiolli and M.\ Ruzzi,
	arXiv:1106.2500.

\bibitem{PosWig}
	D.\ Gross, 
	J.\ Math.\ Phys.\ {\bf 47}, 122107 (2006).
						
\bibitem{CV}
	J.\ Eisert and M.\ B.\ Plenio,
	Int.\ J.\ Quant.\ Inf.\ {\bf 1}, 479 (2003);
	C.\ Weedbrook, S.\ Pirandola, R.\ Garcia-Patron, N.\ J.\ Cerf, T.\ C.\ Ralph, J.\ H.\ Shapiro, and S.\ Lloyd,
	Rev.\ Mod.\ Phys.\ {\bf 84}, 621 (2012).
	
\bibitem{Hudson}
	R.\ L.\ Hudson, Rep.\ Math.\ Phys.\ {\bf 6}, 249 (1974).	
	
\bibitem{Werner}
	T.\ Broecker and R.\ F.\ Werner, J.\ Math.\ Phys.\ {\bf 36}, 62
	(1995);
	A.\ Mandilara, E.\ Karpov, and N.\ J.\ Cerf,
	Phys.\ Rev.\ A {\bf 79}, 062302 (2009).
		
\bibitem{Walgate}	
	J.\ Walgate, A.\ J.\ Short, L.\ Hardy, and V.\ Vedral, 
	Phys.\ Rev.\ Lett.\
	{\bf 85}, 4972 (2000).

\bibitem{BriegelEncoding}
	M.\ Van den Nest, W.\ Duer, A.\ Miyake, and H.\ J.\ Briegel,
	 New J.\ Phys.\ {\bf 9}, 204 (2007).


	
\bibitem{Brandao}
	F.\ G.\ S.\ L.\ Brandao and M.\ B.\ Plenio,
	Nature Phys.\ {\bf 4}, 873 (2008).
	
\bibitem{Oppenheim}
	F.\ G.\ S.\ L.\ Brandao, M.\ Horodecki, J.\ Oppenheim, J.\ M.\ Renes, and R.\ W.\ Spekkens,
	arXiv:1111.3882.		
		
\bibitem{New}
	V.\ Veitch, C.\ Ferrie, D.\ Gross, and
	J.\ Emerson, arXiv:1201.1256, version 4 of Aug.\ 20, 2012.
		
\bibitem{New2}
	V.\ Veitch, C.\ Ferrie, N.\ Wiebe, and
	J.\ Emerson, in preparation (2012).
			
\end{thebibliography}
\end{document}